\documentclass[12pt]{article}
\usepackage{amssymb}

\begin{document}
\title{\bf Plane Symmetric Solutions in $f(R)$ Gravity}

\author{M. Sharif \thanks{msharif@math.pu.edu.pk} and M. Farasat
Shamir \thanks{frasat@hotmail.com}\\\\
Department of Mathematics, University of the Punjab,\\
Quaid-e-Azam Campus, Lahore-54590, Pakistan.}

\date{}

\maketitle
\begin{abstract}
The modified theories of gravity, especially the $f(R)$ theory, have
attracted much attention in recent years. In this context, we
explore static plane symmetric vacuum solutions using the metric
approach of this theory. The field equations are solved using the
assumption of constant scalar curvature which may be zero or
non-zero. We have found a total of three plane symmetric solutions.
The correspondence of these solutions with the well-known solutions
in General Relativity is given.
\end{abstract}


\section{Introduction}

It is a reality that general theory of relativity (GR) has resolved
many problems since its birth. However, there are some issues which
are still problematic. For example, the localization of energy is
still an irritating issue. Also, there are some other problems in
astrophysics and cosmology like the issues of dark energy and dark
matter where we experience severe theoretical difficulties. It has
been found that $96$ percent energy of the universe contains dark
energy and dark matter ($76$ percent dark energy and $20$ percent
dark matter) \cite{1}. Dark matter is basically an unknown form of
matter which has the same properties as ordinary matter but cannot
be detected in the laboratory.

Singularity is another important issue in GR. The singularity
theorem shows that occurrence of the spacetime singularity is a
general feature of any cosmological model under some reasonable
conditions. It may be possible to avoid these undesired
singularities in the context of alternative theories. For example,
it has been shown \cite{28} that there does not exist any
cosmological singularity by considering higher order curvature
terms. Thus there is a need of modified or generalized theories to
tackle these sort of problems. The $f(R)$ theory of gravity is one
of the modified theories which may help to resolve such issues.

The $f(R)$ actions were first studied by Weyl \cite{3} and Eddington
\cite{4} in $1919$ and $1922$ respectively. Later, Buchdahl \cite{5}
studied these actions rigorously in the context of non-singular
oscillating cosmologies. Jakubiec and Kijowski \cite{6} worked on
theories of gravitation with non-linear Lagrangian. They proved that
any theory of gravitation with a non-linear Lagrangian depending on
the Ricci tensor was equivalent to the Einstein theory of
gravitation interacting with additional matter fields. Recent
literature \cite{25}-\cite{21} shows keen interest in exploring
different issues in $f(R)$ theories of gravity. Sotiriou \cite{25}
investigated the relationship between $f(R)$ theory of gravity and
scalar tensor theory. Amendola et al. \cite{7} derived the
conditions under which dark energy $f(R)$ models are cosmologically
viable. Hu and Sawicki \cite{8} verified that the choices of $f(R)$
models have stable high-curvature limits and well-behaved
cosmological solutions with a proper era of matter domination only
if $d^{2}f/dR^{2}>0$.

The most widely explored exact solutions in $f(R)$ gravity are the
spherically symmetric solutions. Multam$\ddot{a}$ki and Vilja
\cite{20} studied spherically symmetric vacuum solutions in this
theory. They found that the whole set of the field equations in
$f(R)$ gravity gave exactly the Schwarzschild de Sitter metric. The
same authors \cite{21} investigated the perfect fluid solutions and
showed that pressure and density did not uniquely determine $f(R)$.
It was found that matching the outside Schwarzschild de Sitter
metric to the metric inside the mass distribution led to additional
constraints that severely limited the allowed fluid configurations.
Capozziello et al. \cite{22} explored spherically symmetric
solutions of $f(R)$ theories of gravity via the Noether symmetry
approach. Hollenstein and Lobo \cite{24} analyzed the exact
solutions of static spherically symmetric spacetimes in $f(R)$
modified theories of gravity coupled to non-linear electrodynamics.

It is interesting, at least from theoretical point of view, to
consider other exact solutions of the field equations in $f(R)$
gravity. Recently, Azadi et al. \cite{23} have studied cylindrically
symmetric vacuum solutions in metric $f(R)$ theories of gravity.
Momeni and Gholizade \cite{024} extended this work to a more general
cylindrically symmetric solution. It would be interesting to extend
this analysis to plane symmetric vacuum solutions.

In this paper, we focuss our attention to investigate the exact
solutions of static plane symmetric (for a class of plane symmetric
spacetims for which the coefficient of $dx^2=1$) vacuum solutions in
$f(R)$ theories of gravity using metric approach. In particular, we
have found three solutions by using the assumption of constant
scalar curvature. The paper is organized as follows: In section
\textbf{2}, we give a brief introduction about the field equations
in the context of $f(R)$ gravity. Section \textbf{3} is used to find
plane symmetric solutions and also the constant curvature solutions.
In the last section, we summarize and conclude the results.

\section{Some Basics of $f(R)$ Gravity }

There are mainly two approaches in $f(R)$ theory of gravity. The
first is called "metric approach" in which the connection is the
Levi-Civita connection and the variation of the action is done
with respect to the metric. The second approach is the "Palatini
formalism" in which the connection and the metric are considered
independent of each other and the variation is done for the two
parameters independently. Here we deal the problem with the metric
approach.

The $f(R)$ theory of gravity is basically the modification or
generalization of general theory of relativity. The action for
$f(R)$ gravity is given by \cite{20}
\begin{equation}\label{1}
S=\int\sqrt{-g}(\frac{1}{16\pi{G}}f(R)+L_{m})d^4x.
\end{equation}
Here $f(R)$ is a general function of the Ricci scalar and $L_{m}$ is
the matter Lagrangian. It may be observed that this action is
obtained by just replacing $R$ with $f(R)$ in the standard
Einstein-Hilbert action. The corresponding field equations are found
by varying the action with respect to the metric $g_{\mu\nu}$
\begin{equation}\label{2}
F(R)R_{\mu\nu}-\frac{1}{2}f(R)g_{\mu\nu}-\nabla_{\mu}
\nabla_{\nu}F(R)+g_{\mu\nu}\Box F(R)=\kappa T_{\mu\nu},
\end{equation}
where
\begin{equation}\label{3}
F(R)\equiv df(R)/dR,\quad\Box\equiv\nabla^{\mu}\nabla_{\mu}
\end{equation}
with $\nabla_{\mu}$ the covariant derivative and $T_{\mu\nu}$ is the
standard matter energy-momentum tensor derived from the Lagrangian
$L_m$. These are the fourth order partial differential equations in
the metric tensor. The fourth order is due to the last two terms on
the left hand side of the equation. If we take $f(R)=R$, these
equations reduce to the field equations of GR.

Now contracting the field equations, it follows that
\begin{equation}\label{4}
F(R)R-2f(R)+3\Box F(R)=\kappa T.
\end{equation}
In vacuum, this reduces to
\begin{equation}\label{5}
F(R)R-2f(R)+3\Box F(R)=0.
\end{equation}
This gives a relationship between $f(R)$ and $F(R)$ which may be
used to simplify the field equations and to evaluate $f(R)$. It is
obvious from this equation that any metric with constant scalar
curvature, say $R=R_{0}$, is a solution of the contracted equation
(\ref{5}) as long as the following equation holds
\begin{equation}\label{6}
F(R_{0})R_{0}-2f(R_{0})=0.
\end{equation}
This is called constant curvature condition. Moreover, if we
differentiate Eq.(\ref{5}) with respect to $R$, we obtain
\begin{equation}
F'(R)R-R'F(R)+3(\Box F(R))'= 0.
\end{equation}
which gives a consistency relation for $F(R)$.

\section{Plane Symmetric Vacuum Solutions}

Here we shall find plane symmetric static solutions of the field
equations in $f(R)$ gravity. For the sake of simplicity, we take the
vacuum field equations and also use constant scalar curvature
$(R=constant)$. In the following, we obtain three particular
solutions of the static plane symmetric spacetimes with these
conditions in $f(R)$ gravity.

\subsection{Plane Symmetric Spacetimes}

We consider the general static plane symmetric spacetime given by
\begin{equation}\label{32}
ds^{2}=A(x)dt^{2}-C(x)dx^{2}-B(x)(dy^{2}+dz^{2}).
\end{equation}
For the sake of simplicity, we take $C(x)=1$ so that the above
spacetime takes the form
\begin{equation}\label{33}
ds^{2}=A(x)dt^{2}-dx^{2}-B(x)(dy^{2}+dz^{2}).
\end{equation}
The corresponding Ricci scalar becomes
\begin{equation}\label{34}
R=\frac{1}{2}[\frac{2A''}{A}-(\frac{A'}{A})^2
+\frac{2A'B'}{AB}+\frac{4B''}{B}-(\frac{B'}{B})^2],
\end{equation}
where prime represents derivative with respect to $x$. Using
Eq.(\ref{5}), it follows that
\begin{equation}\label{9}
f(R)=\frac{3\Box F(R)+F(R)R}{2}.
\end{equation}
Inserting this value of $f(R)$ in the vacuum field equations, we
obtain
\begin{equation}\label{10}
\frac{F(R)R_{\mu\nu}-\nabla_{\mu}\nabla_{\nu}F(R)}{g_{\mu\nu}}
=\frac{F(R)R-\Box F(R)}{4}.
\end{equation}
Since the metric (\ref{33}) depends only on $x$, one can view
Eq.(\ref{10}) as the set of differential equations for $F(x)$, $A$
and $B$. It follows from Eq.(\ref{10}) that the combination
\begin{equation}\label{11}
A_{\mu}\equiv\frac{F(R)R_{\mu\mu}-\nabla_{\mu}\nabla_{\mu}
F(R)}{g_{\mu\mu}},
\end{equation}
is independent of the index $\mu$ and hence $A_{\mu}-A_{\nu}=0$
for all $\mu$ and $\nu$. Thus $A_{0}-A_{1}=0$ gives
\begin{equation}\label{35}
[\frac{A'B'}{AB}+(\frac{B'}{B})^2-\frac{2B''}{B}]F-\frac{1}{2A}(A'F'+2F'')=0.
\end{equation}
Also, $A_{0}-A_{2}=0$ yields
\begin{equation}\label{36}
[\frac{2A''}{A}-(\frac{A'}{A})^2+\frac{A'B'}{AB}-
\frac{2B''}{B}]F-\frac{1}{2}(\frac{A'}{A}-\frac{B'}{B})F'=0.
\end{equation}
Thus we get two non-linear differential equations with three
unknowns namely $A,~B$ and $F$. The solution of these equations
could not be found straightforwardly. However, we can find a
solution using the assumption of constant curvature.

\subsection{Constant Curvature Solutions}

For constant curvature solution, say for $R=R_{0}$, we have
\begin{equation}\label{15}
F'(R_{0})=0=F''(R_{0}).
\end{equation}
Using this condition, Eqs.(\ref{35}) and (\ref{36}) reduce to
\begin{eqnarray} \label{37}
\frac{A'B'}{AB}+(\frac{B'}{B})^2-\frac{2B''}{B}=0,\\\label{38}
\frac{2A''}{A}-(\frac{A'}{A})^2+\frac{A'B'}{AB}- \frac{2B''}{B}=0.
\end{eqnarray}
These equations are solved by the power law assumption, i.e.,
$A\propto x^{m}$ and $B\propto x^{n}$, where $m$ and $n$ are any
real numbers. Thus we use $A=k_1x^{m}$ and $B=k_2x^{n}$, where
$k_1$ and $k_2$ are constants of proportionality. It follows that
\begin{equation}\label{41}
\quad m=-\frac{2}{3}, \quad n=\frac{4}{3}
\end{equation}
and hence the solution becomes
\begin{equation}\label{42}
ds^{2}=k_1x^{-\frac{2}{3}}dt^{2}-dx^{2}-k_2x^{\frac{4}{3}}(dy^{2}+dz^{2}).
\end{equation}
It can be shown that these values of $m$ and $n$ lead to $R=0$. We
re-define the parameters, i.e., $\sqrt{k_1}~t\longrightarrow
\tilde{t},~\sqrt{k_2}~y\longrightarrow \tilde{y}$ and
$\sqrt{k_2}~z\longrightarrow \tilde{z}$, the above metric takes the
form
\begin{equation}\label{42a}
ds^{2}=x^{-\frac{2}{3}}d\tilde{t}^{2}-dx^{2}-x^{\frac{4}{3}}(d\tilde{y}^{2}+d\tilde{z}^{2})
\end{equation}
which is exactly the same as the well-known Taub's metric \cite{30}.

Now we assume $A(x)=e^{2\mu({x})}$ and $B(x)=e^{2\lambda({x})}$ so
that the spacetime (\ref{33}) takes the form
\begin{equation}\label{43}
ds^{2}=e^{2\mu({x})}dt^{2}-dx^{2}-e^{2\lambda({x})}(dy^{2}+dz^{2}).
\end{equation}
The corresponding Ricci scalar is given by
\begin{equation}\label{44}
R=2\mu''+2{\mu'}^2+4\mu'\lambda'+4\lambda''+6{\lambda'}^{2}.
\end{equation}
Using Eq.(\ref{11}), $A_{0}-A_{1}=0$ and $A_{0}-A_{2}=0$
respectively yield
\begin{eqnarray}\label{46}
2(\lambda''+\lambda'^{2}-\mu'\lambda')F-\mu'F'+F''=0, \\
\label{47}
(\mu''-\lambda''+\mu'^{2}-2\lambda'^{2}+\mu'\lambda')F+(\mu'-\lambda')F'=0.
\end{eqnarray}

For constant curvature solutions, the above equations reduce to
\begin{eqnarray}\label{48}
\lambda''+\lambda'^{2}-\mu'\lambda'=0,\\\label{49}
\mu''-\lambda''+\mu'^{2}-2\lambda'^{2}+\mu'\lambda'=0.
\end{eqnarray}
Equation (\ref{48}) can be written as
\begin{equation}\label{50}
\lambda'(\frac{\lambda''}{\lambda'}+\lambda'-\mu')=0
\end{equation}
which leads to the following two cases:
\begin{eqnarray*}
I.\quad \lambda'=0,\quad
II.\quad\frac{\lambda''}{\lambda'}+\lambda'-\mu'=0.
\end{eqnarray*}
Now we solve the field equations for these two cases.

\subsection*{Case I}

It follows from the case I that
\begin{equation}\label{51}
\lambda=a,
\end{equation}
where $a$ is an integration constant. Inserting this value in
Eq.(\ref{49}) and integrating the resulting equation, we obtain
\begin{equation}\label{55}
\mu=\ln(bx+bc),
\end{equation}
where $b$ and $c$ are integration constants. Thus the metric
(\ref{43}) becomes
\begin{equation}\label{56}
ds^{2}=(bx+bc)^2dt^{2}-dx^{2}-e^{2a}(dy^{2}+dz^{2}).
\end{equation}
The corresponding scalar curvature is
\begin{equation}\label{57}
R=0.
\end{equation}
It is mentioned here that the metric (\ref{56}) is a solution only
for those $f(R)$ functions which are linear superposition of $R^m$.
For instance, $f(R)=R+aR^2$ could be the right choice and
$f(R)=R-(-1)^{n-1}\frac{a}{R^{n}}+(-1)^{m-1}bR^{m}$ cannot be used
in this case. The reason is that the function of Ricci scalar
becomes undefined for $R=0$. This solution corresponds to the
self-similar solution of the infinite kind for the parallel dust
case \cite{31}.

\subsection*{Case II}

The second case yields
\begin{equation}\label{58}
\mu'=\lambda'+\frac{\lambda''}{\lambda'}.
\end{equation}
Integrating this equation, we obtain
\begin{equation}\label{59}
\mu=\lambda+\ln\lambda'+d.
\end{equation}
Using the assumption of constant scalar curvature, it follows that
\begin{equation}\label{61}
2\frac{\lambda'''}{\lambda'}+14\lambda''+12{\lambda'}^2=constant
\end{equation}
which is a third order non-linear differential equation. The general
solution of this equation seems to be difficult. However, a special
choice, out of a larger set of possible solutions is that
$\lambda(x)$ is a linear function of $x$, i.e., $\lambda(x)=fx+g$,
where $f$ and $g$ are arbitrary constants. Consequently, the metric
(\ref{43}) takes the form
\begin{equation}\label{62}
ds^{2}=e^{2(fx+\bar{g})}dt^{2}-dx^{2}-e^{2(fx+g)}(dy^{2}+dz^{2}),
\end{equation}
where $\bar{g}=g+\ln f+d$. The corresponding Ricci scalar reduces
to
\begin{equation}\label{63}
R=12f^2.
\end{equation}
Now re-defining $e^{\bar{g}}t\longrightarrow
\tilde{t},~e^{g}y\longrightarrow \tilde{y}$ and
$e^{g}z\longrightarrow \tilde{z}$, it follows that
\begin{equation}\label{62a}
ds^{2}=e^{2fx}(d\tilde{t}^{2}-d\tilde{y}^{2}-d\tilde{z}^{2})-dx^{2}.
\end{equation}
This corresponds to the well-known anti-de Sitter spacetime
\cite{32} in GR. Thus we have found a total of three static plane
symmetric solutions with the assumption of constant scalar curvature
which may be zero or non-zero in $f(R)$ gravity.

\section{Summary and Conclusion}

The purpose of this paper is to study the plane symmetric solutions
in $f(R)$ theories of gravity. We have used the metric approach of
this theory to study the field equations. Since the field equations
in this theory are highly non-linear and complicated, it is very
difficult to solve them analytically without any assumption. We have
developed an important condition, given by Eq.(\ref{6}), for
constant scalar curvature.

We can assume the function $F$ arbitrarily to solve these equations
but this gives fourth order highly non-linear differential
equations. The assumption of constant curvature (may be zero or
non-zero) is found to be the most suitable and we can get some
solution for constant scalar curvature. However, it is not always
guaranteed that solution would be possible using this assumption.
Using this assumption, we have found a total of three static plane
symmetric solutions. It is mentioned here that out of these three
solutions, two are exactly similar to Taub's solution and anti
deSitter spacetime while the third solution corresponds to the
self-similar solution already available in the literature. The
physical relevance of these solutions is very much obvious.

It has been shown \cite{33} that dark energy and dust matter phases
can be achieved by the exact solution derived from $f(R)$
cosmological model. The dark energy is considered as one of the
causes for expansion of universe. In a recent paper \cite{34}, we
have discussed the expansion of universe in the context of metric
$f(R)$ gravity. Thus we expect that such solutions may provide a
gateway towards the solution of dark energy and dark matter
problems. We would like to mention here that most of the work in
$f(R)$ gravity has been done for vacuum static cases. It would be
worthwhile to investigate solutions for non-static and non-vacuum
cases.

\end{document}